\documentclass[graybox, nosecnum]{svmult}

\usepackage{titlesec}
\usepackage{mathptmx}       
\usepackage{helvet}         
\usepackage{courier}        
\usepackage{type1cm}        
\usepackage{makeidx}         
\usepackage{graphicx}        
\usepackage{multicol}        
\usepackage[bottom]{footmisc}
\usepackage{hyperref}        
\usepackage{soul}            
\hypersetup{colorlinks=true,urlcolor=blue}
\usepackage[square,numbers]{natbib}
\makeindex             

\begin{document}
\title*{Mission Design for the TAIJI misson and Structure Formation in Early Universe}
\author{Xuefei Gong, Shengnian Xu,Shanquan Gui, Shuanglin Huang and Yun-Kau Lau \thanks{Corresponding author}}
\institute{Xuefei Gong \at 
 Institute of Applied Mathematics, Academy of Mathematics and System Science, Chinese Academy of Science, Beijing, China, 100190. {\email  xfgong@amss.ac.cn}
 \and 
 Shengnian Xu \at
 Institute of Applied Mathematics, Academy of Mathematics and System Science, Chinese Academy of Science, Beijing, China, 100190. {\email   snxu@geichina.org}
\and
Shanquan Gui \at  School of Physical Science and Technology, Lanzhou University, Lanzhou, 730000, China; Department of Astronomy, School of Physics and Astronomy, Shanghai Jiao Tong University, Shanghai, 200240,  China. \email{ guishq20@sjtu.edu.cn.}
\and Shuanglin Huang \at Institute of Applied Mathematics, Academy of Mathematics and System Science, Chinese Academy of Science, Beijing, China, 100190; University of Chinese Academy of Sciences, Beijing 100049, China. \email{hslynn@amss.ac.cn.}
\and Yun-Kau Lau \at  Morningside Center of Mathematics and Institute of Applied Mathematics, Academy of Mathematics and System Science, Chinese Academy of Science, Beijing, China, 100190.\\ \email{lau@amss.ac.cn.}}
%
%
\maketitle
\abstract{
Gravitational wave detection in space promises to open a new window in astronomy to study the strong field dynamics of  gravitational physics in astrophysics and cosmology.
The present article is an extract of a  report on a feasibility study of gravitational wave detection in space, commissioned by the National Space Science Center, Chinese Academy of Sciences almost a decade ago. The objective of the study was to
explore various possible mission options to detect gravitational waves in space alternative to
that of the (e)LISA mission concept and look into the requirements on the  technological fronts.
On the basis of relative merits and balance between science and technological feasibility, a set of representative mission options were studied and
in the end  a mission design was recommended as the starting point for research and development in the Chinese Academy of Sciences. The mission design was eventually adopted by the current TAIJI mission as the baseline parameters for the project. Subject to technological constraints, the baseline parameters of the TAIJI mission
 were designed in such a way to optimise the  capability of a spaceborne gravitational wave detector to probe high redshift light seed, intermediate mass black holes and thereby shed important light on the structure formation in early Universe.}

\section{Keywords}
Gravitational wave detection in space; Mission design; Intermediate mass black hole binaries at high redshift Universe.

\section{\textit{1. Introduction}}

In the year 2011, the general relativity group of the Morningside Center of Mathematics was commissioned by the National Space Science Center, Chinese Academy of Sciences to conduct a feasibilty study on the detection of gravitational wave in space, as a project in the second phase of the pioneering (Xiandao) program in space. At then, the objective of the study was to
explore various possible mission options to detect gravitational waves in space alternative to
that of the (e)LISA mission concept and look into the requirements on the  technological fronts.
Though some preliminary study was made in the first phase of the pioneering  program \cite{01}, a more complete and systematic study was needed to assess the feasibility for China to undertake further research and development in this direction, in tandem with Europe and US at that time.

A report was submitted at the end of the two year study to the National Space Center in which certain mission design was recommended as a starting point of the development. This mission design was later adopted by the current TAIJI mission as the baseline parameters for the project. In the intervening years,
certain parts of the report were made public both in Chinese and English \cite{02,020}.  The core of the report is concerned with science objectives and it was  translated and published in English \cite{03}. However the English translation left a lot to be desired and obscured the real content of the report. It is the aim of this article to put together the mission design and the primary   science drivers of the TAIJI mission design recommended in the report in a more concise way in English, in the hope that  the content of the report will be accessible to
 a boarder spectrum of readership. At the same time it serves to put on record the early phase of development of gravitational wave detection in space in China.

The outline of the article may be described as follows.  Section 2 will be concerned with a catalogue of the gravitational wave sources at  the frequency window $10^{-4} - 0.1 Hz$. In section 3 we will retrace the line of the thoughts leading eventually to the mission design of the TAIJI mission. In section 4, we will present the simulation of the cosmic growth history of supermassive black holes  and give event rate estimate for the detection of intermediate mass black holes at high redshift Universe. Section 5 is concerned with some simple event rate estimate on the detection of intermediate mass ration inspirals in dense star clusters. Some concluding remarks will be made in section 6 to conclude this article.

\section{2. A survey of gravitational wave sources.}

In comparison with  ground-based
detection, the characteristic mass and energy scale of gravitational wave sources in the detection of  gravitational
wave in space are generally much larger.
Prospective  gravitational wave sources in the frequency window $10^{-4}Hz$ to $0.1Hz$
 are displayed in Table 1.

\begin{table}[]
    \renewcommand\tabcolsep{5.0pt}
    \caption{Expected gravitational wave sources within the frequency window $10^{-4}- 0.1 Hz$.}
    \vspace{10pt}
    \centering
    \begin{tabular}{p{3.5cm}p{8cm}}
        \hline
        Gravitational wave sources  & Brief description of source and the significance of its detection in astronomy and fundamental physics \\

        \hline
        \vspace{5pt}
        Compact binaries (binary white dwarf and  neutron star systems). &
        \vspace{5pt}
        Tens of millions of compact binaries in the Milky Way and beyond. Binaries with precisely measured parameters will be used to calibrate the instruments of a gravitational wave detector and stellar dynamics. The unresolved signals generate a foreground that
        hampers the detection of other gravitational wave sources. At the same time, the foreground provides important information
        on
        the galactic structure of the Milky Way, stellar evolution history and formation mechanism. \\

           \vspace{5pt}

       Intermediate mass black hole binaries at high redshifts. &  \vspace{5pt}
       Intermediate mass black holes descended from the gravitational collapse of heavy Pop III stars, believed to be the seed black holes of the supermassive black holes in galactic centers at the present epoch. The detection will give clues on the galaxy-black hole coevolution and distinguish various models and mechanisms for the cosmic growth  of supermassive black holes and galaxies. \\

        \vspace{5pt}
        Supermassive black holes merger&\vspace{5pt}
        The coalsecence of supermassive black holes that accompany galaxy mergers. The detection provides new way to understand the dynamics of galaxy mergers.
        The high signal to noise ratios also means that it is a good candidate for multi-messenger astronomy and give precision measurement on the redshift-luminosity relation for cosmic evolution.
       \\
        \vspace{5pt}
       Extreme mass ratio inspirals.  &
        \vspace{5pt}
        Compact celestial bodies in the
        galactic centers captured by the supermassive black holes
        or   stars generated by the formation and fast evolution of massive stars
 in the accretion disk surrounding the supermassive black hole. The detection presents a new way to
probe galactic dynamics. The precision measurement of black hole parameters enables us to study spacetime structure
in strong gravity field regime, check on
 the no-hair theorem of black holes  and
distinguish  alternative gravity theories.\\

        \vspace{5pt}
        Intermediate mass ratio inspiral in dense star clusters &
        \vspace{5pt}
        Inspiral systems formed by compact celestial bodies
        and intermediate-mass black holes in the center of dense star cluster.
        Provide unequivocal evidence for  the existence of
        intermediate-mass black holes and reveal the dynamics of star cluster and the formation mechanism for
       intermediate-mass black holes. \\

        \vspace{5pt}
        Gravitational waves in early Universe &
        \vspace{5pt}
        Gravitational waves generated by the big bang, inflation and possibly other cosmological processes like for instance the first order electroweak phase transition.  \\

        \vspace{5pt}
       Burst signals. &
        \vspace{5pt}
        Hyperbolic encounters of black holes or celestial bodies and possibly from unknown sources.  \\

        \vspace{5pt}
        Unmodelled sources. &
        \vspace{5pt}
        New physics and new astronomy \\
        \vspace{5pt}&\\

        \hline

    \end{tabular}
    \label{table1}
\end{table}

\subsection{2.1 Compact Binary Star Systems.}
Through  optical and X-ray observations, a number of compact binary star systems with inspiral period shorter than one hour in the Milky Way were identified,
including some binary
systems with mass-transfer taking place. These binaries are confirmed gravitational wave sources at the millihertz band. It is expected that, with the increase in precision in our observations, more binaries of this kind \cite{26,27,28,49}  will be discovered.  As the
astrophysical parameters and positions
of these systems may be measured precisely, the gravitational waves  generated by these sources may be predicted in a very accurate way \cite{bender-hils, farmer-phinney}.
These sources, termed verification binaries, may be used in a reverse way as a standard candle to calibrate a gravitational wave detector in space.

  Though
weak in strength in the gravitational wave amplitude in comparsion with those generated by binary black hole mergers, due to the
abundance in numbers (up to millions) and much shorter distance from a spaceborne detector, currently planned spaceborne detector will detect gravitational wave signals generated by millions of compact binaries in and outside of the Milky Way. Data analysis will enable us to identify signals from
around ten thousand individual compact binaries and determine
 the astrophysical parameters of these wave sources, such as the mass, luminosity
distance, position, and the orbital inclination of the binary stars. The rest unresolved signals will become a foreground (confusion) noise around the $mHz$ band and hamper the detection of other gravitational wave sources. At the same time, the foreground signal will be modulated by the annual motion of a spaceborne detector with respect to the orientation of the Galactic disk. The strength, spectral
shape and annual variation of the foreground   may then also be used to constrain and probe galactic structure, stellar evolution history and formation mechanism
which are otherwise difficult to probe by conventional electromagnetic astronomy.

\subsection{2.2 Binary Black Hole Mergers}
These are the majority of gravitational wave sources for detection in space and holds the promise of unravelling important
and significant information concerning astrophysical and cosmological physics. Black hole binaries may be classified in accordance with the mass ratios between the
two black holes undergoing inspiral and coalescence processes.

\subsubsection{2.2.1 Intermediate mass black holes binaries at high redshift Universe}

Astronomical Observations indicate that  supermassive black holes are present in the centers of both
normal and active galaxies, and that the velocity dispersion of stars in the galactic bulge is
closely correlated with the mass of the supermassive black hole in the galactic center, i.e,
the $M-\sigma$ relation \cite{2,3,4} . The $M-\sigma$ relation reveals that a co-evolution history possibly
exists between the galaxy and its central black hole.
 Quasars observed by the SDSS at redshifts larger than 6
powered by the accretion of supermassive black holes with  $\sim 10^ 9 M_\odot$ in mass require
that the seeds of supermassive black holes should form in the earlier period of the Universe \cite{5,6,7}. This gives important constraints to
the formation of supermassive black holes and the occurrence of the high-contrast density
fluctuations in early universe.

Observational results of the cosmic background radiation indicate that energy fluctuations
of the fractional order of $10^{-5}$ exist on the isotropic background after the  recombination era at $z\sim 1000$. The gravitational (Jeans) instability induced collapse and aggregation in these
high-density places of the early Universe, and the first generation dark matter halos formed
at around $z=20\sim30$, in which the gas finally experienced cooling, fragmentation, and
agglomerating of $H_2 $ molecules and the Pop III ( first generation) stars as a result \cite{8,9}.
A lack of metal abundance nor efficient cooling mechanism mean that the Pop III stars are very massive. The first light  from these stars ended the dark age era
of the Universe, while its energetic UV component ionised the fragile neutral $H_2$ molecules and heralded the dawn of
 reionization.
With  zero metal abundance, the stellar wind loss caused by
radiation may be neglected, the lost mass in the final stage of collapse is very small. Numerical
simulations indicate that the Pop III stars of $100\sim 260M_\odot$  disappear directly through
the supernova explosions due to electron-positron pair instability, while the stellar bodies with
heavier mass  collapsed to form the black holes with  mass greater than $\sim 150 M_\odot$ \cite{10}.
This falls in the range of intermediate mass black holes  which is one possible origin of the seed black holes for supermassive black holes observed at the present epoch.
Alternatively, gas clouds may directly collapse to form the seed black holes with
mass heavier than $10^4 M_\odot$. Through an effective angular momentum transfer, the collapse
of stellar cluster nuclei may also form  seed black holes of $10^3 \sim 10^4 M_{\odot}$ \cite{17,18}.

According to the hierarchical growth model of cold  dark matter, the  supermassive black holes located at  center of a galaxy at the present epoch evolved from
density fluctuations before re-ionization: the small-mass sub-galactic structures formed first,
 then merged with each other continuously to form bigger and bigger structures \cite{10,11,12}
In this process, if black holes existed in the galactic centers, these black holes would
experience the continuous accretion and merger processes of the dark matter halos and
host galaxies, and formed the currently observed massive black holes
at different redshifts \cite{11}-\cite{19}.

A detailed understanding of the formation and growth of supermassive black holes is of clear cosmological as well
as astrophysical significance \cite{190}. It represents in part of  our endeavour to understand galaxy evolution and structure formation of the Universe.
Electromagnetic astronomy enables us to probe the dynamics of a galaxy at redshifts around 6 or a little higher. It is expected that
 the prospective launch of the JWST  telescope will extend our horizon to higher redshifts in the infrared red regime of electromagnetic waves.
 At high redshifts beyond the reach of electromagnetic astronomy,   gravitational waves detection provides a unique way to probe the growth of black holes
 and galaxies through probing the merger of intermediate mass black holes binaries descended from the Pop III stars.   Gravitational waves may be considered as an
 entirely new
window to observe the galaxy-black hole co-evolution process and  obtain  important information
regarding the history of structure evolution beginning from the end of the dark age era \cite{20,21}.

\subsubsection{2.2.2 Supermassive black hole mergers.}

In the merger of two galaxies, provided there are supermassive black holes residing in the
the centers of both galaxies, the two supermassive black holes will be dragged into the newly
formed galactic center due to the dynamic friction with the surrounding stars, and gradually a distance is  reached within which
gravitational interaction becomes dominant and a bound state is formed.  Through the mutual three-body interaction with the surrounding small
celestial bodies or the friction with  gas, the potential energy is reduced and the orbit
is further shrunk. When the binary black hole system transfers their gravitational energy
and angular momentum to the surrounding stars through the interactions with them, the
surrounding stars acquire  higher energy and angular momentum and become capable of escaping from the galactic center.
When the stellar environment is deficient in gas and the galactic structure is spherical symmetric,
there is then a lack of dynamical drag that drives further the merger of two black holes (i.e., the final parsec problem). Current study indicates that after merger
the structures of galaxies themselves are not spherical symmetric and the the stellar orbits
are aspherical. There are enough stars capable of getting into the vicinity of the
binary black holes, so that the two black holes can keep their closeness by ejecting stars,
and the distance between the black holes further reduces; finally,  gravitational waves
generated by the gravitational interaction of the binary black holes begins to play a dominant role and brings away the
energy and angular momentum, ending up finally coalescence of the binary black holes.

 In the case when the stellar environment is gas rich, after the formation of binary black holes, a common accretion disk
for the binary black holes forms from the surrounding gas. The binary
black holes further accrete matter, transfer the angular momentum to surrounding stars and finally forms a single black hole through the emission of gravitational waves \cite{13,14,15,17,56}.
Many complicated astrophysical factors and processes are involved, such as  gas
accretion, dynamic interaction with the small stars, structure of the accretion disk,
accretion efficiency, and  gravitational recoil of the final merger, etc. Detection
of gravitational waves provides an entirely new way to the understanding of these  physical processes
\cite{21,57,58}

The signal-noise ratio for the detection of the merger of two supermassive black holes is in general very large.
The gravitational wave signals generated in the final merging process  is detectable even in the time domain. Through the accumulation in time
sufficient signal to noise ratio (for example, one month or one year before merger), certain  astrophysical
parameters like for instance the luminosity distance, orbital plane and possibly other parameters of the binary supermassive black hole merging system may be precisely measured and these otherwise will be difficult to obtain by electromagnetic means.  Combined with the information of redshifts from measurements in
electromagnetic wave astronomy,  the binary supermassive black hole merger wave sources will
become a standard siren with large signal-noise ratio and redshift range (in analogy
to the standard candle of electromagnetic wave astronomy). This will enable us to determine precisely the redshift-luminosity
relation and other cosmic evolution information \cite{20,21}.

\subsubsection{2.2.3 Extreme mass ratio inspirals.}

In the center of a galaxy, through dynamical capture of a compact celestial body (stellar mass black hole, neutron star or a white
 dwarf) by the supermassive black hole located at the galactic center or possibly certain companion stars generated by the formation and fast evolution of massive stars
 in the accretion disk surrounding the supermassive black hole, a binary system, termed extreme mass ratio inspiral (EMRI), is formed with  mass ratios $\sim 10^6$  or bigger between the supermassive black hole
 and the compact celestial body. EMRIs  are an important gravitational wave source
for detection in space around the $mHz$ regime \cite{62,63,64}.
Though the event rate is highly uncertain,  the general expectation is that EMRIs should be present in a galactic center.
A long duration of observation is needed to accumulate the EMRI signal to noise ratio  to a level above the threshold for detection.

The detection of
gravitational waves from  EMRIs  provides a new window to study and explore the galactic center region which
otherwise would be difficult to resolve by means of electromagnetic observations.
Through probing the dynamics of EMRIs near the galactic center, we will be able to
measure the mass, spin and higher order
multipole moments of a supermassive black hole and the detectable range extends far beyond the local Universe. The detection  is also a way to confirm the existence of a supermassive black hole in the center of an inactive galaxy in the $(10^5 \sim 10^7 M_{\odot})$ regime
  and enable us to do precision measurements on black hole parameters.

The power of gravitational radiation generated by an EMRI system is rather
weak and the orbital evolution of the compact celestial body  is very slow. The adiabatic evolution of  Kerr black hole
geodesics may be used to describe the orbital dynamics generated by  gravitational waves.
Though the periodically averaged frequency variation is generally unremarkable, due to the
existence of eccentricity and a number of precession effects, the gravitational waves emitted by
an EMRI system has rich short-period frequency variations and the phases with complex
structures \cite{68}-\cite{72}. As a  gravitational wave source, an EMRI system is characterised
by 17 parameters. It is a great challenge to  handle the adiabatic and multiscale evolution of the spacetime structure and the corresponding geodesics.
The data processing in a large and complex parametric space together with a weak and long time wave train will also be a problem we have to contend with. The mass,
angular momentum, mass quadrupole moment of a supermassive black hole  retrieved
 from the  gravitational wave signals of
EMRIs \cite{73,74} also presents a way to check on the no-hair theorem of black holes \cite{68, 76}  and
to distinguish  alternative gravity theories.

\subsubsection{2.2.4 Intermediate mass ratio inspirals.}

The observations of ultraluminous X-ray sources (ULXs) and stellar dynamics in star clusters
suggest the possible existence of intermediate-mass black holes of $10^2 \sim 10^4 M_{\odot} $ in a dense star clusters.
Observations indicate that ultraluminous compact sources are very common in our local Universe, the  well known ones
are M82-X1, M15, NGC3628, and the Centaurus $\omega$ etc. \cite{70, 78, 79, 80}. Generally, the ULX sources locate at
a certain distance from the  centers of galaxies, which means that their mass is not
very large (does not exceed $10^5 M_{\odot}$). It is found that some ULX sources are surrounded by
 gaseous loops illuminated by X-rays, this means that at least the X-rays emitted by a part
of ULX sources are not highly concentrated in the line of sight connecting the source to the Earth.
If the energy released by the ULX sources is isotropic and the luminosity does not exceed the Eddington limit, it may be deduced that
the mass for these X-ray sources is in the range of $10^2 \sim 10^4 M_{\odot}$. A strong quasi-periodic
light variability has been detected for some ULX sources. The accretion  by an
intermediate-mass black holes at the center of a star cluster  is one possible explanation for these ULX sources.

 N body  simulations of compact star clusters also suggests that intermediate-mass black holes can be produced by the fast collision and merging of stellar
mass black holes in the central regions of star clusters, or by the repeated aggregation of massive stars.  Some simulations even
suggest that multi intermediate mass black holes may form in young star clusters \cite{81}-\cite{85}. In the star clusters
containing intermediate-mass black holes, the compact celestial bodies, such as the stellar mass
 black holes, neutron stars, etc., can be captured by the intermediate-mass black hole
through the dynamic processes of two-body exchange or hierachical three-body interaction to form a binary system with mass ratios of a few  tens to one thousand \cite{83}-\cite{88},
which is termed the intermediate mass ratio inspiral (IMRI). Further, when a star cluster
with an intermediate-mass black hole evolves toward the galactic nucleus, it will  gradually be
torn up by the tidal force, the remnant intermediate-mass black hole and the supermassive black
hole in the galactic center will form an IMRI on a bigger mass scale \cite{87}.

   At present,   the theory
that  an intermediate-mass black hole in the center of a dense star  cluster provides a plausible explanation of the ULX observations. on, very strong
There are however many other possible
explanations for the observations. Conventional electromagnetic astronomy often is not able to provide good resolution of the center of a cluster.  However, gravitational waves can be violently emitted in the
final inspiral stage of  IMRI systems. The gravitational wave detection of IMRIs in dense star clusters will give
equivocal evidence for the existence of intermediate-mass black holes in dense stellar environment. At the same time,
the detection
provides an entirely new avenue to probe the dynamics of star clusters and addresses important astrophysical problems like
 the formation mechanism of intermediate-mass
black holes, the initial mass-function and dynamical evolution history of star clusters etc.

The characteristic frequencies of gravitational waves emitted by IMRIs in a dense star clusters fall
into the $0.01\sim 1Hz$ regime. To optimise detection of these sources, baseline parameters of a spaceborne
detector is required to tune to this frequency range as the most sensitive regime of a detector.
At the same time, better sensitivity in laser metrology is also needed to compensate for the weak signal strength
due to smaller intermedate mass scale \cite{87}-\cite{90}.

\subsubsection{2.2.5 Gravitational waves  from  early Universe}

Gravitational waves generated during the big bang and subsequent inflation era provide the earliest clue in our
quest for
understanding of the origin and creation of our Universe. The primary science objectives of The BBO (Big Bang Observer) \cite{100,101}
 and the DECIGO (Deci-Hertz Gravitational wave Observatory) \cite{102,103,104} project
 are to detect directly the primordial
gravitational waves  from the early universe. However these projects pose extremely demanding requirements on the technological fronts.
With the foreseeable development in technologies in the next two decades, it is more sensible to be content with giving an upper bound on this
gravitational wave background. There are many competing theories and models which seek to describe the dynamics soon after the big bang, among these are the Big Bang-inflation process, the first-order electroweak phase transition,
the extra-dimensional dynamics predicted by the superstring theory, the cosmic string network,
\cite{20,99} etc.
The current mission design will be able to impose meaningful constraints on the parameter space of these models.

\subsubsection{2.2.6 Burst signals.}

Burst signals are difficult to forecast their waveforms and foresee their detection at this stage.
 These signals may be
originated from the short-distance capture/grazing among celestial bodies, or from  cosmological process like the
 breaking up of a cosmic string and possibly some new and unmodelled  astrophysical or even quantum gravitational processes.

 \section{3. Mission design}

The starting point of the study is to explore the feasibility of detection in the band gap  from $0.1Hz$ to $10Hz$  between LISA and LIGO. To this end,
 it is natural to consider in the mission design  to shorten the armlength of the interferometer and increase the precision of the laser metrology by suppressing shot noise and enlarge the diameter of the telescopes. This happens to overlap with the
ALIA mission concept proposed by Peter Bender \cite{bender1, bender2} which
is conceivably in many ways the simplest adaptation of the LISA mission to a frequency band
one order of magnitude higher than that of LISA, with the promise of mapping out
mass and spin distribution of light seed black holes at  high redshifts.

However, when the key technologies of  the ALIA mission was further looked at, the
sub-picometer interferometry requirement in the laser metrology part poses a major obstacle on
the technological side of the mission. With a view that China will have a reasonable chance to
realise the mission in the next few decades and to minimise possible risks in future research and development  of the
key technologies, further relaxation of the baseline parameters to a more realistic level seems to be a natural step to take.

Based on the relative merits between science and technological viability, a number of mission options from both scientific and  technological perspective
are carefully studied. Mainly due to foreseeable technological limitations in laser interferometry in the next two decades,
in the end a compromise between scientific significance and technological feasibility is reached and the following baseline parameters were chosen for further studies given in Table 2. Apart from the detection of gravitational wave sources in the $10^{-4}-0.1Hz$ frequency window, the primary science driver is set to  probe  high redshifts intermedate mass black hole binaries, with a view to understand structure formation in early Universe and galaxy-black hole co-evolution. In doing so, the mission will become a part of an astronomy program, working closely with future infrared red astronomy and radio astronomy programs in China to explore high redshift Universe after the dark age era.  The baseline parameters are subject to minor variations, in particular the position noise may be relaxed further to 10pm or more, yet the science of the mission is still worth pursuing.

\begin{table}[]
    \renewcommand\tabcolsep{5.0pt}
    \caption{}
    \vspace{10pt}
    \centering
    \begin{tabular}{ccccc}
        \hline
        arm length          & telescope        & laser           & 1-way position              & acceleration noise \\
        ($m$)               & diameter($m$)    &power($W$)         & noise($pm\cdot Hz^{-1/2}$)  & ($m\cdot s^{-2}\cdot Hz^{-1/2}$) \\
        \hline
        $3\times10^9$        &0.45$\sim$0.6       &2              &5$\sim$8                & $3\times10^{-15}(>0.1mHz)$ \\
        $5\times10^8$(ALIA)  &1.0                 &30             &0.1                     & $3\times10^{-16}(>1mHz)$ \\
        $5\times10^9$(LISA)  &0.4                 &2              &18                      & $3\times10^{-15}(>0.1mHz)$ \\
        $1\times10^9$(eLISA) &0.2                 &2              &11                      & $3\times10^{-15}(>0.1mHz)$ \\
        \hline
    \end{tabular}
    \label{table-11}
\end{table}

For reference purpose, the baseline design parameters of ALIA, LISA/eLISA are also given in
Table 2. The relevant sensitivity curves are displayed in Figure 1. Apart from the instrumental
noises, confusion noise generated by both galactic and extra-galactic compact binaries are also
taken into consideration. Relevant confusion levels are converted from estimations by Hils and
Bender \cite{bender-hils} and  Farmer and Phinney \cite{farmer-phinney}.

\begin{figure}[h]
\centering
\includegraphics[scale=0.25]{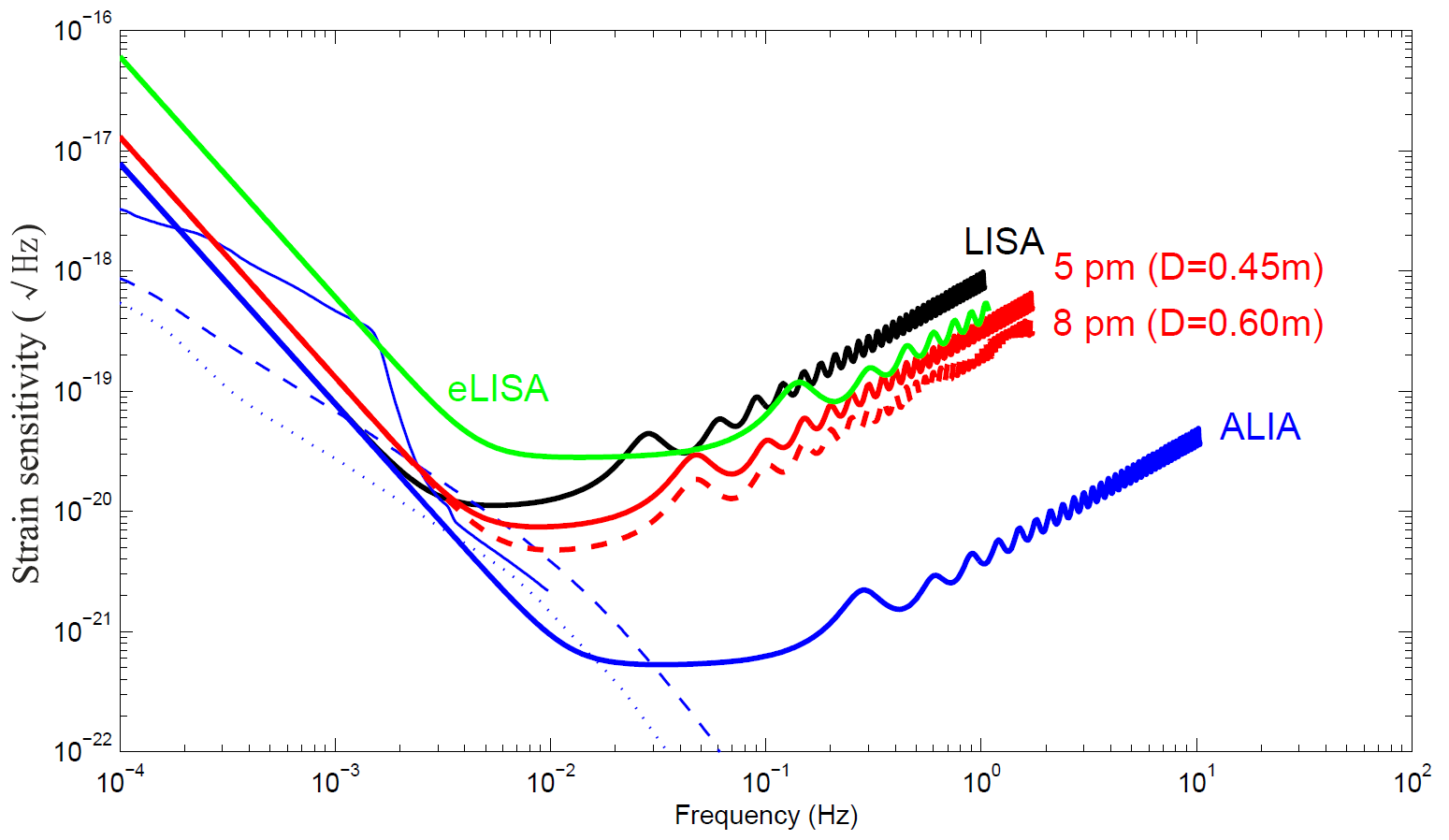}
\caption{Sensitivity curves of mission designs with different choice of baseline parameters, with ALIA, LISA and eLISA included for the purpose of comparsion.}
\label{pic1}
\end{figure}

For black hole binaries with mass ratio $1:4$, typical of what one would expect from
hierarchical black hole growth at high redshift, the all angle averaged detection range are plotted
in Figure 2. Apart from galactic confusion noise, upper level (dashed curve) and lower level
(dotted dashed curve) of confusion noise generated by extragalactic compact binaries as those
estimated by \cite{farmer-phinney} are also taken into account.

In calculating the averaged SNR, we have used hybrid waveforms in the frequency domain
with black hole spin not taken into account \cite{ajith1,ajith2}. For one year of observation before merger, the
contributions in SNR due to large spin is indeed negligible according to our calculations. Spin is
relevant only in the parameter estimation stage, which will not be discussed in the present work.
As may be seen from Figure 2, for a given redshift, the proposed mission concept is capable
of detecting lighter black hole binaries in comparsion with eLISA/LISA and thereby provides
better understanding of the hierarchical assembling process in early Universe.

\begin{figure}[h]
\centering
\includegraphics[scale=0.25]{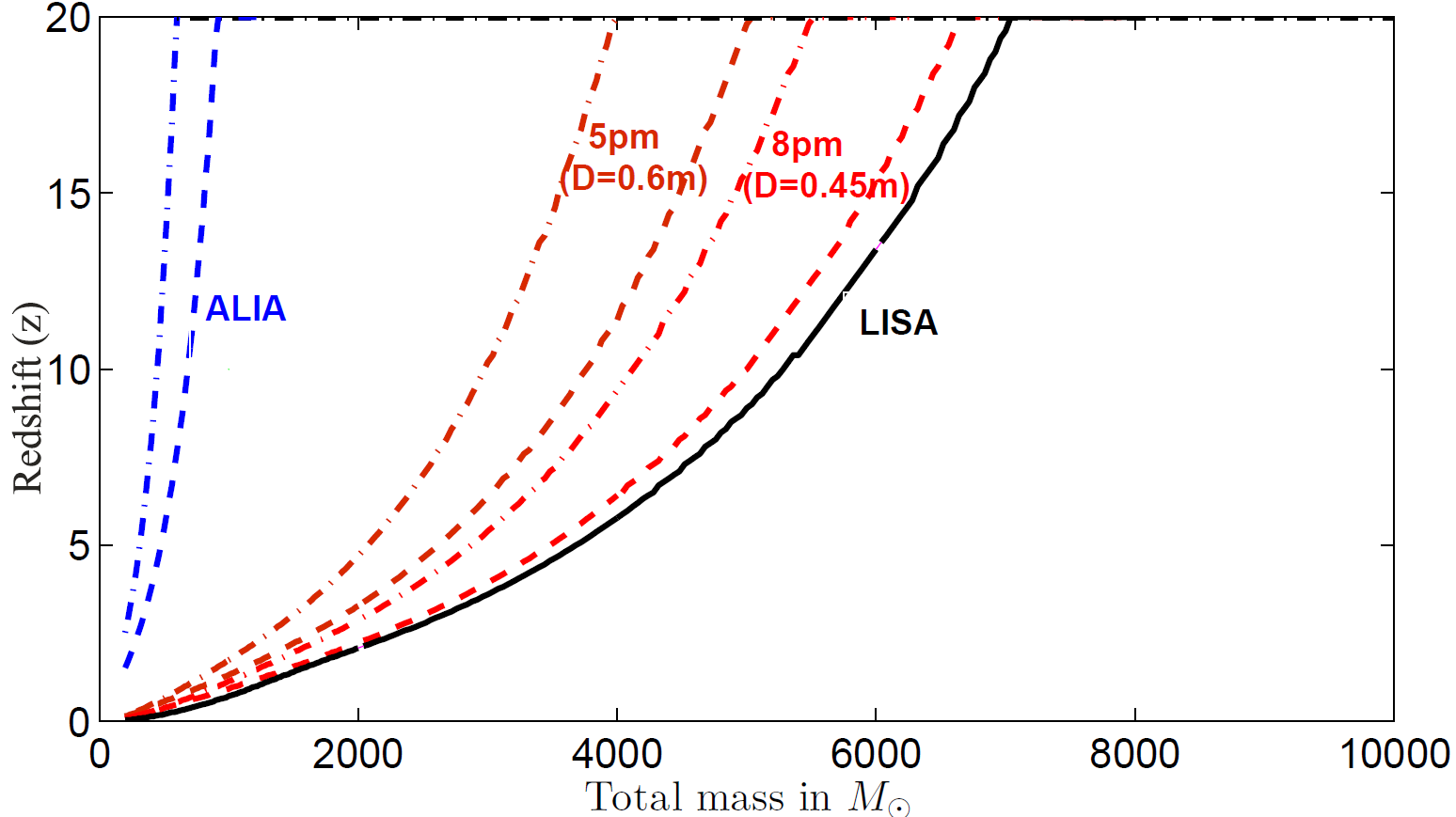}
\caption{All-angle averaged detection range of a single Michelson channel with threshold SNR of 7 for
$1:4$ mass ratio intermediate black hole binaries, one year observation prior to merger. For each mission
option, both upper and lower confusion noise levels (represented by the dashed curve and dotted
dashed curve respectively) due to extragalactic compact binaries are considered.}
\label{pic2}
\end{figure}

\begin{figure}[h]
\centering
\includegraphics[scale=0.25]{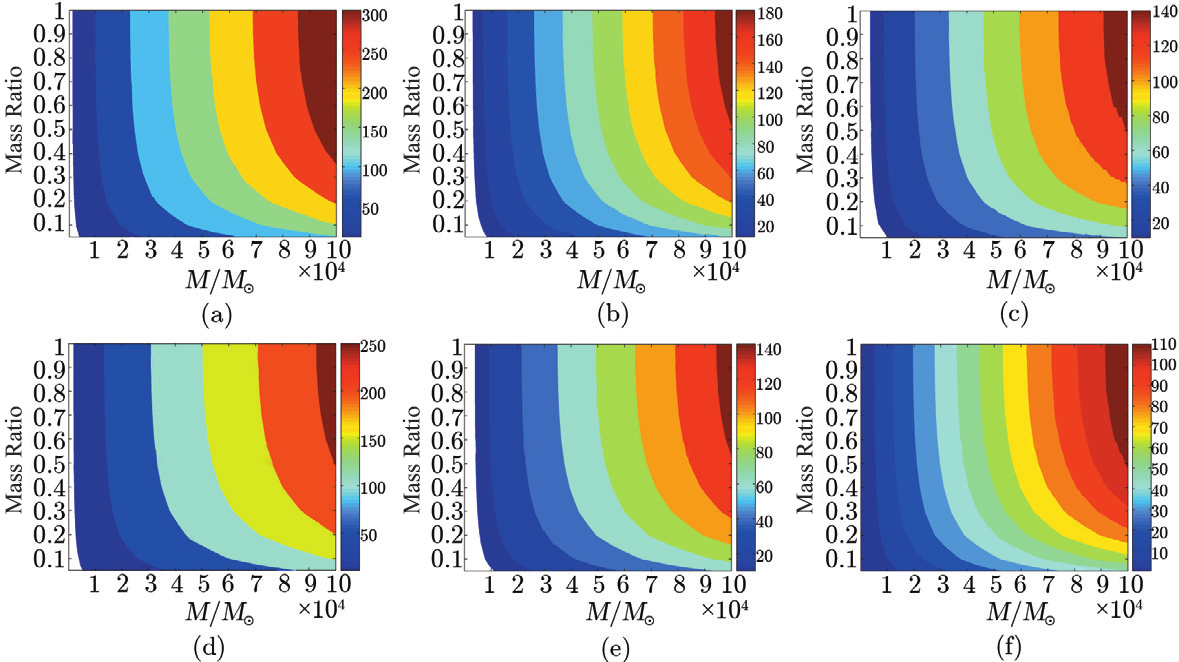}
\caption{The signal-noise ratio contours for detecting IMRIs at different redshifts by a single Michelson channel with  $3\times 10^6km$ armlength and one year observation before merger.  (a) $5pm\cdot Hz^{-1/2}$, $z=2$,
(b) $5pm\cdot Hz^{-1/2}$, $z=6$, (c) $5pm\cdot Hz^{-1/2}$, $z=10$, (d) $8pm\cdot Hz^{-1/2}$, $z=2$,
(e) $8pm\cdot Hz^{-1/2}$, $z=6$, and (f) $8pm\cdot Hz^{-1/2}$ , $z=10$.
Confusion noise considered is at an intermediate level between the upper and lower limit}
\label{fig5}
\end{figure}

Apart from intermediate mass black hole binaries at high redshift, the designed sensitivity at around $0.01Hz$
measurement band means that the instrument is also capable of detecting IMRIs  harboured at globular clusters or dense young star clusters at low redshift
($z<0.6$). See \cite{88} for a further discussion of the capture dynamics of an IMRI in dense star
clusters. Displayed in Fig 3 are the detection ranges of IMRIs with different mass ratios one
year prior to merger. The stellar black hole  is fixed to be $10M_{\odot}$, while
the mass of an intermediate mass black hole is subject to variation in order to generate different mass ratios in the figure.

\begin{figure}[h]
\centering
\includegraphics[scale=0.25]{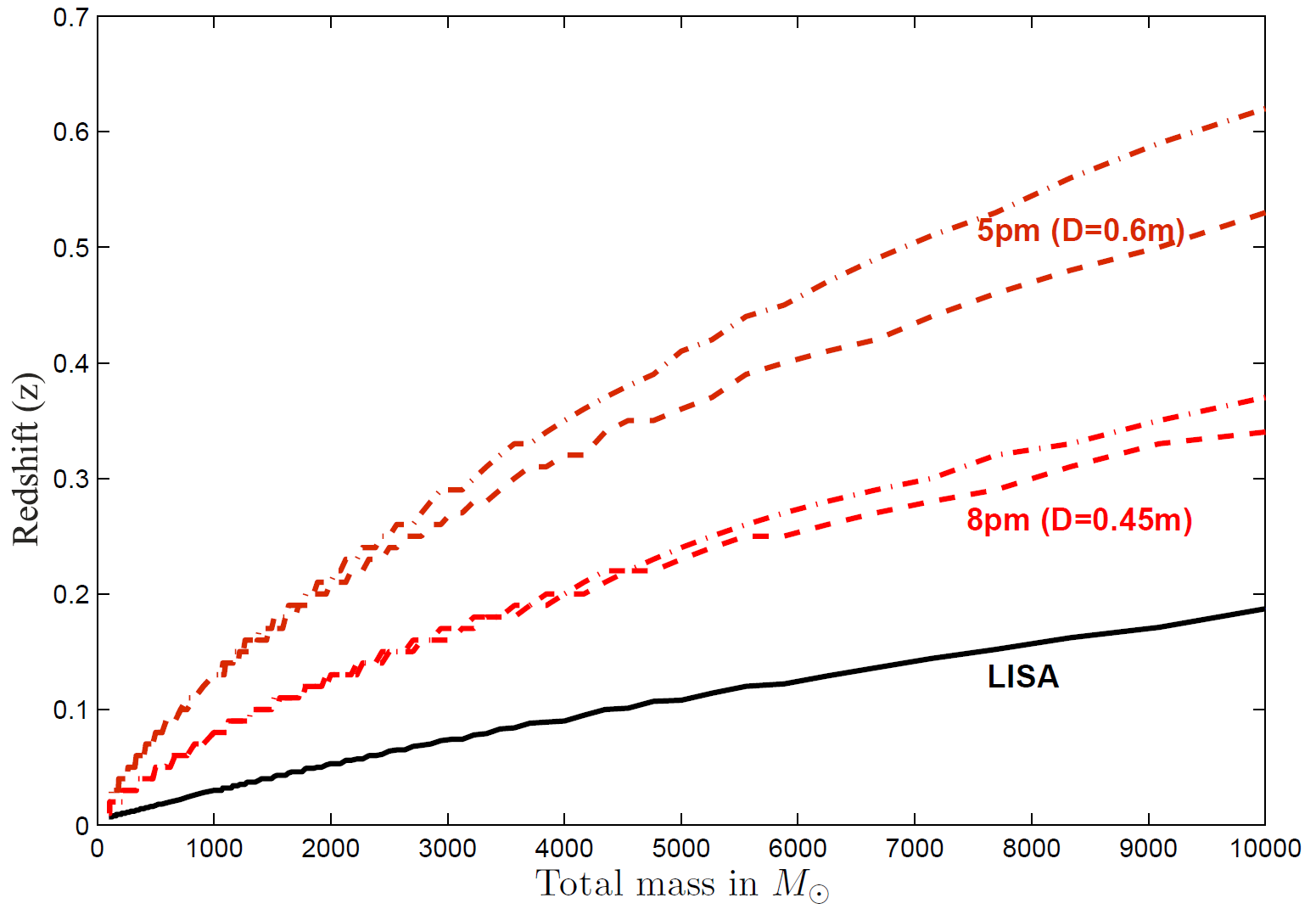}
\caption{All-angle averaged detection range under a single Michelson threshold SNR of 7
for IMRIs with reduced masses of $10M_{\odot}$, one year
observation prior to merger. For each mission option, both upper and lower confusion noise levels
(represented by the dashed curve and dotted dashed curve respectively) due to extragalactic
compact binaries are considered.}
\label{pic3}
\end{figure}

\begin{figure}[h]
\centering
\includegraphics[scale=0.25]{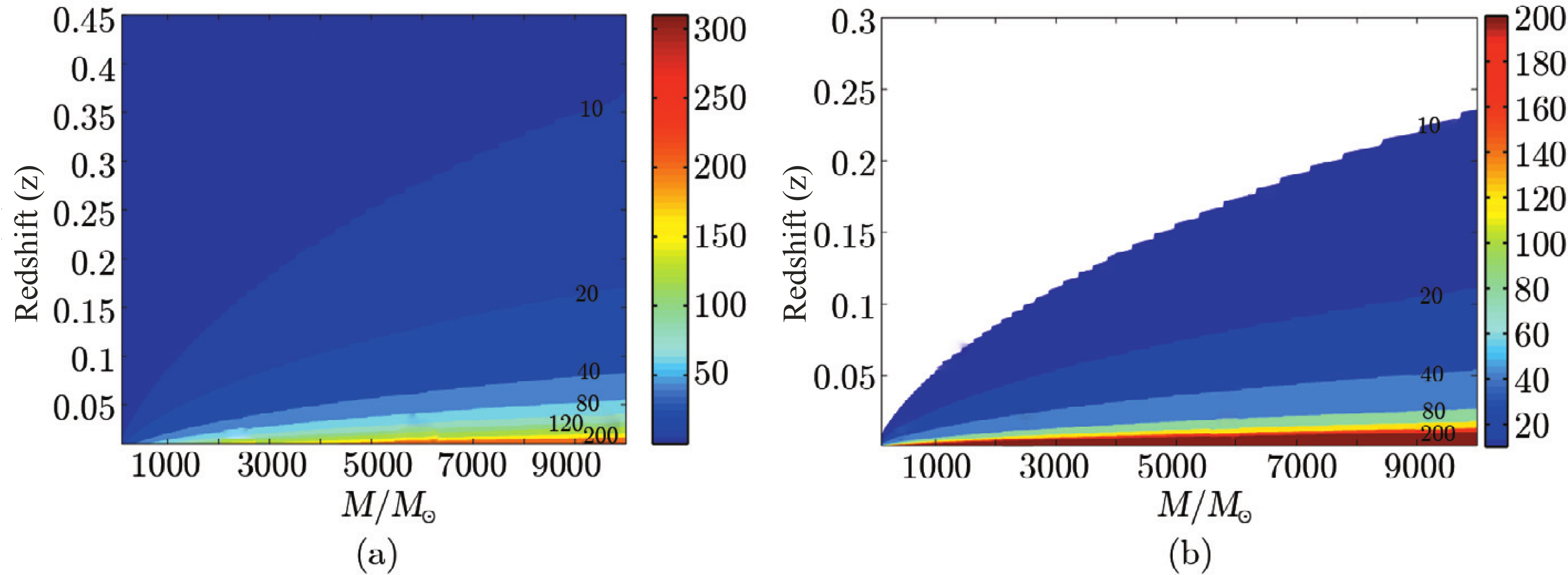}
\caption{The signal-noise ratio contours for detecting IMRIs
    with the reduced mass of $10M_{\odot}$) by a single Michelson channel with armlength of
$3 \times 10^6km$, one year of observation before merger.
    (a) $5pm\cdot Hz^{-1/2}$, and (b) $8pm\cdot Hz^{-1/2}$. The confusion noise considered is at an intermediate level between the
upper and lower limits.}
\label{fig3}
\end{figure}

\section{4. Simulations of  cosmic growth and merger of black holes and event rate estimates}

To understand the detection capability in high redshift Universe  of the mission options given in Table 2,
 we carry out a Monte Carlo simulation of black hole merger histories based
on the EPS formalism and semi-analytical dynamics.

Pop III remnant black holes of $150M_{\odot}$ are placed in $3.5\sigma$ biased halos at $z=20$ with initial
spins of the seeds generated randomly. By prescribing VHM-type dynamics \cite{13,14}, we trace
downwards the black hole merging history. The halo mass ratio criteria for major merger is set
to be greater than $0.1$. Both the prolonged accretion and the chaotic accretion scenario are considered.
Black hole spins coherently evolve through both mergers and accretions processes and their
magnitudes influence strongly the mass-to-energy conversion efficiency. We assume efficient
gaseous alignment of the black holes so that the hardening time is short and only moderate
gravitational radiation recoils take place.  Numerical simulations \cite{khan} suggest that the hardening
and merger time scales remain short even in gas free environment.
In calculations relevant to
GW observations, we assume a threshold SNR of 7 for detection in the sense of single Michelson
interferometer and one year observation prior to merger.

The results are schematically given in
Figure 6 and Figure 7.

\begin{figure}[h]
\centering
\includegraphics[scale=0.30]{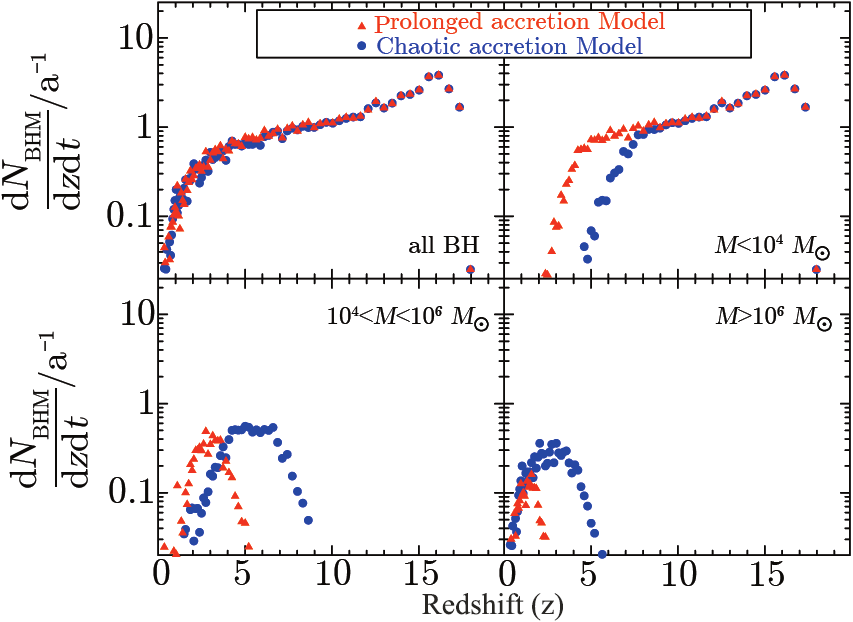}
\caption{Coalescence rate predicted by the Monte Carlo simulations.}
\label{pic4}
\end{figure}

\begin{figure}[h]
\centering
\includegraphics[scale=0.35]{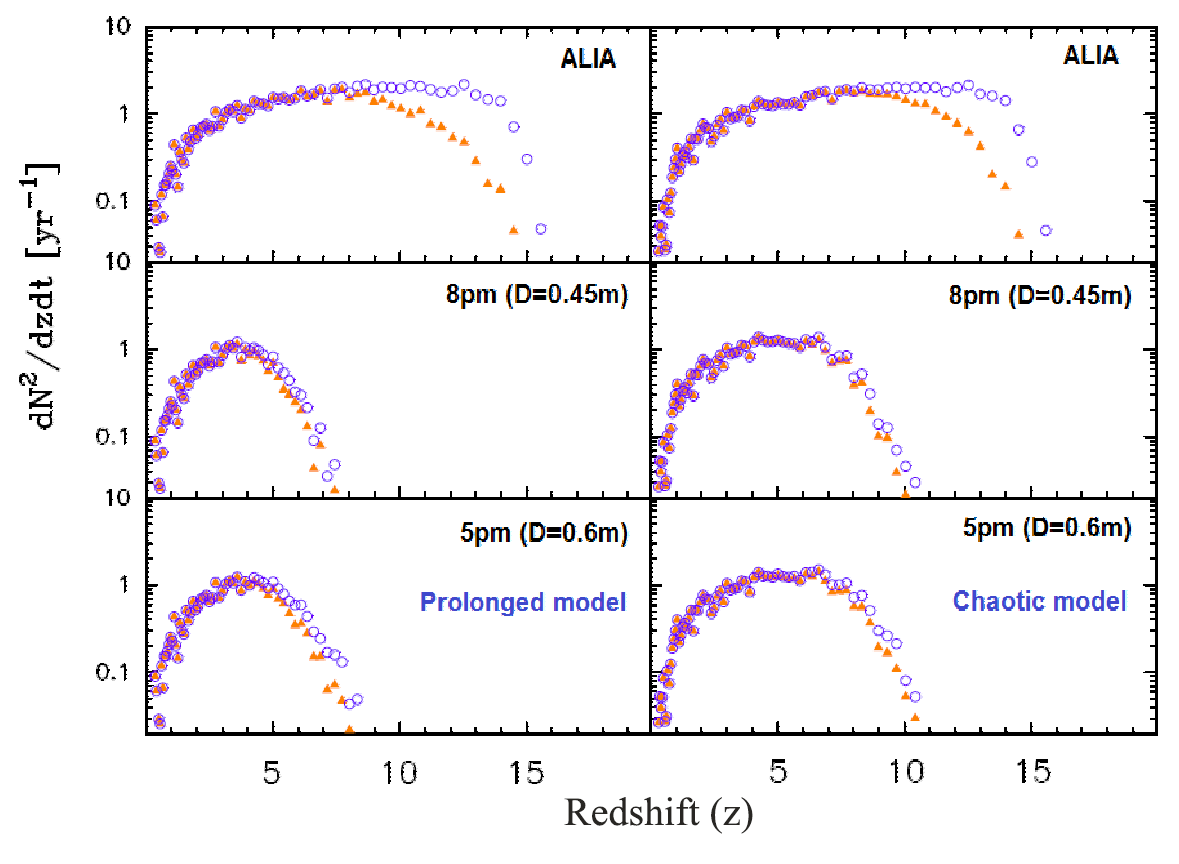}
\caption{Event rates estimates for the mission designs. }
\label{pic5}
\end{figure}

We assess our simulations by fitting the black hole mass functions and luminosity functions
at six almost equally divided successive redshift intervals ranging from $z=0.4$ to $2.1$ (see Figure 8). In the
prolonged accretion scenario, the results deviate from the observational constrains given by
Soltan type argument when going up to redshift $z>1.5$. It may therefore underestimate the
black holes growth rate and perhaps also the coalescence rate. Observationally the existence of
very high redshift ($z>6$) AGNs implies that feed back mechanisms may be very different at
early epoch so that fast growth of the seed black hole could be possible.

\begin{figure}[h]
\centering
\includegraphics[scale=0.22]{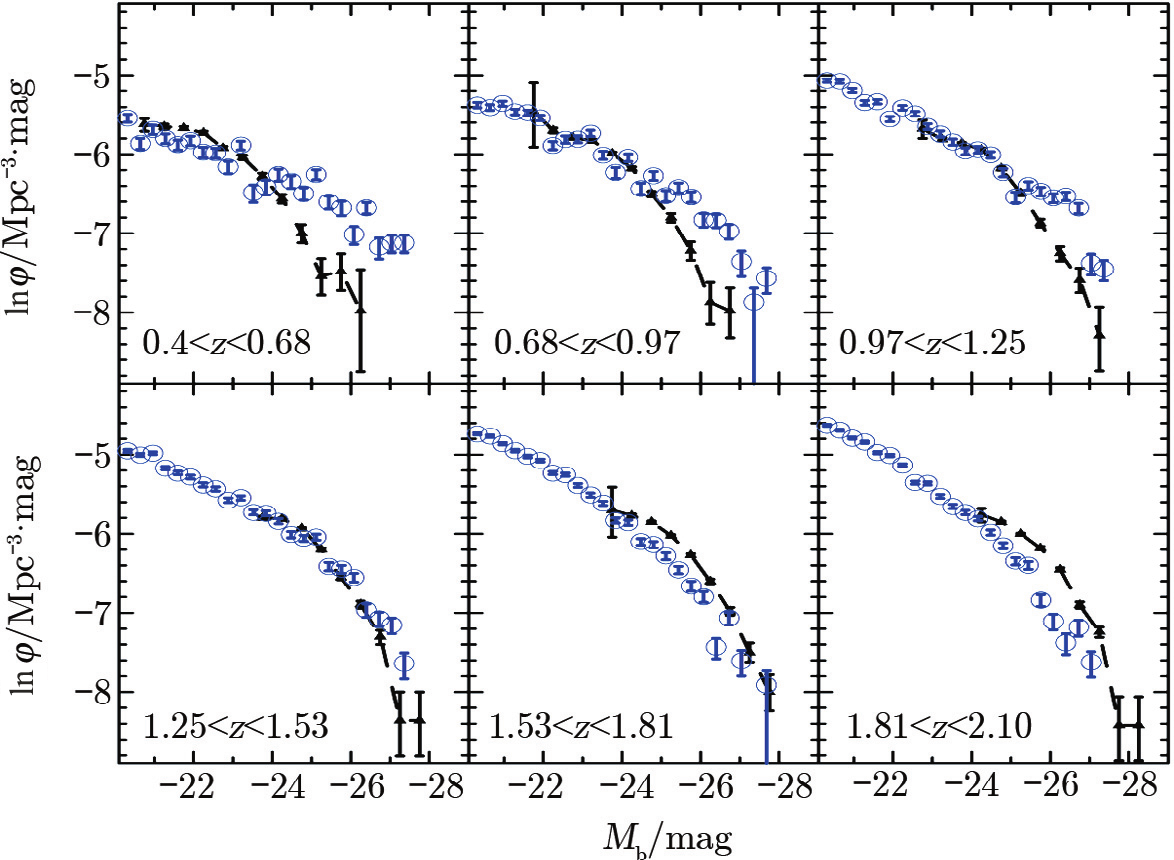}
\caption{The degree of coincidence of the luminosity function of quasars given by the Monte Carlo simulation
with  observed data}
\label{fig6}
\end{figure}

In terms of coalescence rate, our result displayed in Figure 4 is in overall agreement with the
results given by Sesana et al \cite{21,57}
and Arun et al \cite{arun}, though the coalescence counts given
by their simulations are about two or three times higher. It is likely due to various numerical
discrepancies in the simulations. Overall, our black hole mass growth is slower, particularly in
the prolonged accretion scenario. At $z=15$, the total mass of the black hole binaries typically
are still less than $600M_\odot$ in the prolonged model and this may lead to a smaller counts in
detectable sources. Our results are expected to give a very conservative (pessimistic) estimate of
black hole binaries merger event rate.

The astrophysics encapsulated in our simulation represents the state of art understanding of
structural formation after the dark age. Due to our poor understanding of the evolution of the
Universe at this epoch, it is likely that the simulation overlooks many details of the physical
processes involved. The event rate count should be looked upon in a cautious way. Instead of
reading into the precise numbers, it serves as an indication what spaceborne gravitational wave
detector is capable of and in our case, the advantage of setting the most sensitive regime of the
measurement band from a few $mHz$ to $0.01Hz$.

\section{5. Event rate estimate for the detection of IMRIs in dense star clusters}

Consider  the  following scenario:
\cite{84,85,86}:

(1) In the accessible range of the universe for the mission design, the spatial number
density of star clusters is a constant in respect to the volume calculated by the luminosity
distance. It has the same value as that of  the local universe.  For globular  and young
clusters: $n_{GC} \approx 8h^3 \cdot Mpc^{-3}$, $n_{YC} \approx 3h^3 \cdot Mpc^{-3}$,
 $h=H_0/(100km \cdot s^{-1} \cdot Mpc^{-1})$ and we take $h=0.73$. In the event rate estimate in what follows, we consider only the
globular clusters and  assume that the total number density of dense star clusters is given by $n_C \approx 8h^3 \cdot Mpc^{-3}$,
which will give a conservative estimation of the event rate.

(2) The probability of a small compact celestial body being captured by an
intermediate-mass black hole at the center of a cluster is given by
$$\nu(M,\mu,z)\approx 10^{-10}\frac{M}{\mu}a^{-1},$$
i.e., it is directly proportional to the mass of the  intermediate-mass black hole,
and inversely proportional to the reduced mass. This assumption is motivated by the
dynamical analysis on  globular clusters \cite{82}-\cite{85}.

In the event rate calculations, for wave sources with
 non-negligible redshift effect,  a scaling factor $1+z$ is required in principle, but the coalescence
rate itself is just an  of order of magnitude estimate. Further, the accessible range of redshifts considered by the
current mission options  is
quite small. This scaling factor will be neglected here.

(3) In dense star clusters where the intermediate-mass black holes is located,
the mass distribution function of the intermediate-mass black holes is assumed to be
$$f(M)=\frac{f_{tot}}{ln\frac{M_{max}}{M_{min}}}\frac{1}{M}.$$
$M_{{min}}$ and $M_{{max}}$ are respectively the lower and upper limits of the mass-distribution range
of intermediate mass black holes, they are taken respectively as $10^2M_{\odot}$ and $10^4M_{\odot}$, and beyond this
range $f(M)=0$. $f_{tot}$ is the fraction of dense star clusters containing  intermediate-mass
black holes. Its value is highly uncertain and we shall take a conservative estimate that $f_{tot}=0.1$ \cite{84,86}.

Take the observational time of one year before merger,
the reduced mass of an IMRI system is taken to be $10M_{\odot}$  and the threshold value of signal-
noise ratio of a single Michelson detection is taken as 7.
The event rate may then be calculated using the following formula \cite{84,87}:
$$R=\frac{4\pi}{3}\int^{M_{max}}_{M_{min}}[D_L(M,\mu)]^3\nu(M,\mu,z)n_c f(M)dM.$$
and the result is given in Table 3.

\begin{table}[]
    \renewcommand\tabcolsep{10.0pt}
    \caption{Prospective detection rate for IMRIs in globular clusters.}
    \vspace{10pt}
    \centering
      \begin{tabular}{lcc}
        \hline
        Mission option      &Upper level of confusion        &Lower level of confusion\\
        \hline
        ALIA                &$z_c=5 : \sim 8000$             &$\sim 12000$          \\
                            &$z_c=3 : \sim 6000$             &$\sim 7000$           \\
        \hline
        5pm(D=0.6m)              &$\sim 90$         &$\sim 130$          \\
        \hline
        8pm(D=0.45m)             &$\sim 26$         &$\sim 32$          \\
        \hline
        LISA ($5\times 10^6$ km in armlength)              &\multicolumn{2}{c}{$\sim 3$}               \\
        \hline
      \end{tabular}
    \label{table2}
\end{table}

The above event rate estimate is subject to many uncertainties and perhaps we should not
attach too much importance to the precise numbers. Instead, the calculations serves as an
indication of the detection potential of the mission concept as far as IMRIs at low redshifts are
concerned. Further, as event rate goes up as the cubic of the improvement in sensitivity, it also
brings out the advantage of shifting slightly the most sensitive region of the measurement band
to a few hundredth $Hz$, as far as detection of IMRIs is concerned. It should also be remarked that
collision of dense star clusters \cite{88} constitutes a possible intermediate mass black hole binaries gravitational wave sources, while
the inspiral of massive black holes ($\sim10^3M_{\odot}$ to $\sim10^4M_{\odot}$) into the supermassive black hole
at the center of a galaxy is also a promising IMRI source \cite{87, liu}. However, the corresponding
event rates would be difficult to estimate.

\section{Concluding remarks}

Gravitational wave detection in space promises to  open a new window in the quest for understanding of our Universe,
in particular as a new way to probe the formation of galactic structure at early Universe discussed here. With the recent
 revised definition of the LISA mission in which the armlength of the laser interferometer is shortened to 2.5 million kilometers,  modulo some minor variations in the baseline parameters, there is basically no longer any difference
between LISA and TAIJI in terms of the mission definition.  A global effort to realise such a mission seems to be the next natural step forward, though it does not seem possible in view of the current political climate. Still it is an option we should keep in mind, in the hope that when the occasion is right this becomes a realistic step to be taken.

\section*{Acknowledgements}

 We were very grateful to Prof. Shing-Tung Yau, director of the Morningside Center of Mathematics,  for very strong financial backing that enabled us to conduct the study in a  very comprehensive way and thereby laid a solid foundation for the current development in China. His  encouragement also served as a source of inspiration for the study. The feasibility study was conducted with some very generous technical assistance from the Albert Einstein Institute, Hannover and Peter Bender. Our thanks also go to Profs. Wenrui Hu and Shuangnan Zhang for support.


\begin{thebibliography}{}

\bibitem{01}
Gong XF, Xu SN, Bai S, et al. (2011) A scientific case study of an advanced LISA
mission. Classical and Quantum Gravity 28(9):094012.

\bibitem{02}
Gong XF, Lau YK, Xu SN, et al (2015) Descope of the ALIA mission.
In: Journal of Physics: Conference Series (Vol. 610, No. 1, p. 012011). IOP Publishing.

\bibitem{020}
Gong XF, Xu SN, Yuan YF, et al. (2015) Laser interferometric gravitational wave detection
in space and structure formation in the early universe(in Chinese). Progress in Astronomy, 33(1):59-83.

\bibitem{03}
Gong XF, Xu SN, Yuan YF, et al. (2015) Laser interferometric gravitational wave detection
in space and structure formation in the early universe. Chinese Astronomy and Astrophysics, 39(4):411-446.

 \bibitem{26}
Burgay M, D'Amico N, Possenti A, et al. (2003) An
increased estimate of the merger rate of double neutron stars from
observations of a highly relativistic system. Nature 426(6966):531-533.


\bibitem{27}
Lyne AG, Burgay M, Kramer M, et al. (2004) A double-pulsar system:
a rare laboratory for relativistic gravity and plasma
physics. Science 303(5661):1153-1157.


\bibitem{28}
Antoniadis J, Freire PC, Wex N, et al. (2013) A massive
pulsar in a compact relativistic binary. Science 340(6131).

\bibitem{49}
Marsh T (2011) Double white dwarfs and LISA. Classical and Quantum
Gravity 28(9):094019.

\bibitem{50}
Nelemans G, Zwart SP, Verbunt F, Yungelson L (2001) Population
synthesis for double white dwarfs-II. Semi-detached systems: AM CVn
stars. Astronomy \& Astrophysics 368(3):939-949.

\bibitem{bender-hils}
Bender PL, Hils D (1997) Confusion noise level due to galactic and
extragalactic binaries. Classical and Quantum Gravity 14(6):1439.

\bibitem{farmer-phinney}
Farmer AJ, Phinney ES (2003) The gravitational wave background
from cosmological compact binaries. Monthly Notices of the Royal
Astronomical Society 346(4):1197-1214.

\bibitem{2}
Ferrarese L, Merritt D (2000) A fundamental relation between supermassive
black holes and their host galaxies. The Astrophysical Journal
Letters 539(1):L9.

\bibitem{3}
Gebhardt K, Bender R, Bower G, et al. (2000) A relationship
between nuclear black hole mass and galaxy velocity dispersion. The
Astrophysical Journal Letters 539(1):L13.

\bibitem{4}
Ferrarese L (2002) Beyond the bulge: a fundamental relation between
supermassive black holes and dark matter halos. The Astrophysical
Journal 578(1):90.


\bibitem{5}
Fan X, Narayanan VK, Lupton RH, et al. (2001) A
survey of $z>5.8$ quasars in the Sloan Digital Sky Survey. I. Discovery of
three new quasars and the spatial density of luminous quasars at $z\sim 6$.
The Astronomical Journal 122(6):2833.

\bibitem{6}
Fan X, Strauss MA, Schneider DP, et al. (2001) High-redshift
quasars found in Sloan Digital Sky Survey commissioning data.
IV. Luminosity function from the fall equatorial stripe sample. The
Astronomical Journal 121(1):54.

\bibitem{7}
Fan X, Hennawi JF, Richards GT, et al. (2004) A survey
of $z>5.7$ quasars in the Sloan Digital Sky Survey. III. Discovery of five
additional quasars. The Astronomical Journal 128(2):515.

\bibitem{8}
Madau P, Rees MJ (2001) Massive black holes as population III rem-
nants. The Astrophysical Journal Letters 551(1):L27.

\bibitem{9}
Heger A, Woosley SE (2002) The nucleosynthetic signature of
population III. The Astrophysical Journal 567(1):532.

\bibitem{10}
Press WH, Schechter P (1974) Formation of galaxies and clusters of
galaxies by self-similar gravitational condensation. The Astrophysical
Journal 187:425-438.

\bibitem{17}
Koushiappas SM, Zentner AR (2006) Testing models of supermassive
black hole seed formation through gravity waves. The Astrophysical
Journal 639(1):7.


\bibitem{18}
Koushiappas SM, Bullock JS, Dekel A (2004) Massive black hole seeds
from low angular momentum material. Monthly Notices of the Royal
Astronomical Society 354(1):292-304.


\bibitem{11}
Lacey C, Cole S (1993) Merger rates in hierarchical models of
galaxy formation. Monthly Notices of the Royal Astronomical Society
262(3):627-649.

\bibitem{12}
Cole S, Lacey CG, Baugh CM, Frenk CS (2000) Hierarchical galaxy formation.
Monthly Notices of the Royal Astronomical Society 319(1):168-204.


\bibitem{13}
Volonteri M, Haardt F, Madau P (2003) The assembly and merging
history of supermassive black holes in hierarchical models of galaxy
formation. The Astrophysical Journal 582(2):559.



\bibitem{14}
Volonteri M, Madau P, Haardt F (2003) The formation of galaxy stel-
lar cores by the hierarchical merging of supermassive black holes. The
Astrophysical Journal 593(2):661.


\bibitem{15}
Volonteri M, Madau P, Quataert E, Rees MJ (2005) The distribution
and cosmic evolution of massive black hole spins. The Astrophysical
Journal 620(1):69.


\bibitem{16}
Volonteri M, Rees MJ (2005) Rapid growth of high-redshift black holes.
The Astrophysical Journal 633(2):624.




\bibitem{19}
Begelman MC, Volonteri M, Rees MJ (2006) Formation of supermassive
black holes by direct collapse in pre-galactic haloes. Monthly Notices
of the Royal Astronomical Society 370(1):289-298.


\bibitem{190}
National Research Council (2010) New Worlds, New Horizons in Astronomy and
Astrophysics. The National Academies Press, Washington, DC.


\bibitem{20}
Sathyaprakash BS, Schutz BF (2009) Physics, astrophysics and cosmology with gravitational waves.
Living reviews in relativity, 12(1):2.


\bibitem{21}
Sesana A, Volonteri M, Haardt F (2007) The imprint of massive black
hole formation models on the LISA data stream. Monthly Notices of
the Royal Astronomical Society 377(4):1711-1716.


\bibitem{56}
Berti E, Volonteri M (2008) Cosmological black hole spin evolution by
mergers and accretion. The Astrophysical Journal 684(2):822.

\bibitem{57}
Sesana A, Gair J, Berti E, Volonteri M (2011) Reconstructing the
massive black hole cosmic history through gravitational waves. Physical
Review D 83(4):044036.


\bibitem{58}
Gair JR, Sesana A, Berti E, Volonteri M (2011) Constraining properties
of the black hole population using LISA. Classical and Quantum
Gravity 28(9):094018.


\bibitem{62}
Amaro-Seoane P, Gair JR, Freitag M, Miller MC, Mandel I, Cutler
CJ, Babak S (2007) Intermediate and extreme mass-ratio inspirals-
astrophysics, science applications and detection using LISA. Classical
and Quantum Gravity 24(17):R113.

\bibitem{63}
Miller MC, Freitag M, Hamilton DP, Lauburg VM (2005) Binary encounters
with supermassive black holes: zero-eccentricity LISA events.
The Astrophysical Journal Letters 631(2):L117.

\bibitem{64}
Levin Y (2007) Starbursts near supermassive black holes: young stars
in the galactic centre, and gravitational waves in LISA band. Monthly
Notices of the Royal Astronomical Society 374(2):515-524.

\bibitem{68}
Ryan FD (1995) Gravitational waves from the inspiral of a compact
object into a massive, axisymmetric body with arbitrary multipole moments.
Physical Review D 52(10):5707.

\bibitem{69}
Ryan FD (1997) Accuracy of estimating the multipole moments of a
massive body from the gravitational waves of a binary inspiral. Physical
Review D 56(4):1845.

\bibitem{70}
Barack L, Cutler C (2007) Using LISA extreme-mass-ratio inspiral
sources to test off-kerr deviations in the geometry of massive black
holes. Physical Review D 75(4):042003.

\bibitem{71}
Finn LS, Thorne KS (2000) Gravitational waves from a compact star in
a circular, inspiral orbit, in the equatorial plane of a massive, spinning
black hole, as observed by LISA. Physical Review D 62(12):124021.

\bibitem{72}
Drasco S (2006) Strategies for observing extreme mass ratio inspirals.
Classical and Quantum Gravity 23(19):S769.


\bibitem{73}
Cornish NJ (2011) Detection strategies for extreme mass ratio inspirals.
Classical and Quantum Gravity 28(9):094016.

\bibitem{74}
Gair JR, Porter E, Babak S, Barack L (2008) A constrained Metropolis-Hastings
search for EMRIs in the mock LISA data challenge 1b. Classical
and Quantum Gravity 25(18):184030.

\bibitem{76}
Drasco S, Hughes SA (2004) Rotating black hole orbit functionals in
the frequency domain. Physical Review D 69(4):044015.

\bibitem{78}
Matsushita S, Kawabe R, Matsumoto H, Tsuru TG, Kohno K, Morita
KI, Okumura SK, Vila-Vilar\'o B (2000) Formation of a massive black
hole at the center of the superbubble in M82. The Astrophysical Journal
Letters 545(2):L107.

\bibitem{79}
Van der Marel RP (2003) Intermediate-mass black holes in the universe:
A review of formation theories and observational constraints. arXiv
preprint astro-ph/0302101.

\bibitem{80}
Fabbiano G (2005) The hunt for intermediate-mass black holes. Science
307(5709):533-534.

\bibitem{81}
Maccarone TJ, Kundu A, Zepf SE, Rhode KL (2007) A black hole in
a globular cluster. Nature 445(7124):183-185.

\bibitem{82}
Zwart SFP, McMillan SL (2002) The runaway growth of intermediate-mass
black holes in dense star clusters. The Astrophysical Journal 576(2):899.

\bibitem{83}
Miller MC, Hamilton DP (2002) Production of intermediate-mass
black holes in globular clusters. Monthly Notices of the Royal
Astronomical Society 330(1):232-240.

\bibitem{84}
Miller MC (2002) Gravitational radiation from intermediate-mass black
holes. The Astrophysical Journal 581(1):438.

\bibitem{85}
Miller MC, Colbert EJM (2004) Intermediate-mass black holes.
International Journal of Modern Physics D 13(01):1-64.


\bibitem{86}
Miller MC (2005) Probing general relativity with mergers of super-
massive and intermediate-mass black holes. The Astrophysical Journal
618(1):426.

\bibitem{87}
Will CM (2004) On the rate of detectability of intermediate-mass black
hole binaries using LISA. The Astrophysical Journal 611(2):1080.

\bibitem{88}
Konstantinidis S, Amaro-Seoane P, Kokkotas, KD (2013) Investigating the retention
of intermediate-mass black holes in star clusters using N-body simulations.
Astronomy \& Astrophysics, 557:A135.


\bibitem{89}
Mandel I, Gair JR (2009) Can we detect intermediate mass ratio inspirals?
Classical and Quantum Gravity 26(9):094036.

\bibitem{90}
Gair JR, Mandel I, Miller MC, Volonteri M (2011) Exploring intermediate
and massive black-hole binaries with the einstein telescope.
General Relativity and Gravitation 43(2):485-518.

\bibitem{99}
Binetruy P, Bohe A, Caprini C, Dufaux JF (2012) Cosmological backgrounds
of gravitational waves and eLISA/NGO: phase transitions, cos-
mic strings and other sources. Journal of Cosmology and Astroparticle
Physics 2012(06):027.

\bibitem{100}
Phinney S, Bender P, Buchman R, et al. (2004) The
Big Bang Observer: Direct detection of gravitational waves from the
birth of the universe to the present. NASA Mission Concept Study.

\bibitem{101}
Crowder J, Cornish NJ (2005) Beyond LISA: Exploring future gravitational
wave missions. Physical Review D 72(8):083005.

\bibitem{102}
Seto N, Kawamura S, Nakamura T (2001) Possibility of direct
measurement of the acceleration of the universe using 0.1 hz band laser
interferometer gravitational wave antenna in space. Physical Review
Letters 87(22):221103.

\bibitem{103}
Kawamura S, Ando M, Nakamura T, et al. (2008) The Japanese space gravitational wave antenna; DECIGO.
Journal of Physics: Conference Series 120(3):032004.

\bibitem{104}
Ando M, Kawamura S, Seto N, et al. (2010) DECIGO
and DECIGO pathfinder. Classical and Quantum Gravity 27(8):084010.

\bibitem{bender1}
Bender PL (2004) Additional astrophysical objectives for LISA follow-
on missions. Classical and Quantum Gravity 21(5):S1203.


\bibitem{bender2}
Bender PL, Begelman MC (2005) Massive black hole formation and growth.
In: Favata F, et al(eds) ESA SP-588, European Space Agency, p33.


\bibitem{ajith1}
Ajith P, Babak S, Chen Y, et al. (2008) Template
bank for gravitational waveforms from coalescing binary black holes:
Nonspinning binaries. Physical Review D 77(10):104017.

\bibitem{ajith2}
Ajith P, Hannam M, Husa S, et al. (2011) Inspiral-merger-ringdown
waveforms for black-hole binaries with nonprecessing spins.
Physical Review Letters 106(24):241101.


\bibitem{arun}
Arun KG, Babak S, Berti E, et al. (2009). Massive black-hole binary inspirals:
results from the LISA parameter estimation taskforce. Classical and Quantum Gravity, 26(9):094027.

\bibitem{khan}

Khan FM, Holley-Bockelmann K, Berczik P, Just A (2013) Supermassive black hole
binary evolution in axisymmetric galaxies: the final parsec problem is not a problem.
The Astrophysical Journal, 773(2):100.


\bibitem{liu}
Liu FK, Li S, Komossa S (2014) A milliparsec supermassive black hole binary candidate
in the galaxy SDSS J120136.02+300305.5. The Astrophysical Journal, 786(2):103.


\end{thebibliography}
\end{document}